# λ-randomization: multi-dimensional randomized response made easy

Nicolas Ruiz

Universitat Rovira i Virgili
Departament d'Enginyeria Informàtica i Matemàtiques
Av. Països Catalans 26, 43007 Tarragona, Catalonia
nicolas.ruiz@urv.cat

**Abstract.** Randomized response is a popular local anonymization approach that can deliver anonymized multi-dimensional data sets with rigorous privacy guarantees. At the same time, it can ensure validity for exploratory analysis and machine learning tasks as, under fairly general conditions, unbiased estimates of the underlying true distributions can be retrieved. However, and like for many other anonymization techniques, one of the main pitfalls of this approach is the curse of dimensionality. When coping with data sets with many attributes, one quickly runs into unsustainable computational costs for estimating true distributions, as well as a degradation in their accuracies. Relying on new theoretical insights developed in this paper, we propose an approach to multi-dimensional randomized response that avoids these traditional limitations. From simple yet intuitive parameterizations of the randomization matrices that we introduce, we develop a protocol called λ-randomization that entails low computational costs to retrieve estimates of multivariate distributions, and that makes use of solely three simple elements: a set of parameters ranging between 0 and 1 (one per attribute of the data set), the identity matrix, and the all-ones vector. We also present an empirical application to illustrate the proposed protocol.

## 1 Introduction

Data on individual subjects are increasingly gathered and exchanged. By their nature, they provide a rich amount of information that can inform statistical and policy analysis in a meaningful way. However, due to the legal obligations surrounding these data, this wealth of information is often not fully exploited to protect the confidentiality of respondents and to avoid privacy threats. In fact, such requirements shape the dissemination policy of individual data at national and international levels. The issue is how to ensure a sufficient level of data protection to meet releasers' concerns in terms of legal and ethical requirements, while still offering users a reasonable level of information.

Statistical disclosure control, that is anonymization techniques for privacy-preserving data publishing, has a rich history of addressing those issues by providing the analytical apparatus through which the privacy/information trade-off can be assessed and implemented [1]. Among the universe of techniques available, randomized response (RR) has the attractive feature of offering rigorous privacy guarantees (expressible, for example, in terms of differentially private guarantees), while at the same time ensuring data validity for exploratory analysis and machine learning tasks [2]. RR can also be run locally or globally. For the former, each individual anonymizes her data before handing them to a (generally untrusted) data controller, while for the latter the controller collects all the data and performs the randomization, a procedure called PRAM [3].

However, and despite these appealing features, like for many other anonymization techniques RR suffers from dimensionality issues. As we will remind below, applying RR simultaneously to a large set of attributes is practically challenging, while dealing with attributes separately generally entails too much loss in data utility. Additionally, dimensionality in the local approach to RR is an especially thorny issue, as by definition the global data picture is missing.

**Contribution and plan of this paper**

Past contributions have attempted to tackle these issues by leveraging specific distributional properties of the data to be randomized, mainly by taking advantage of data sparsity (e.g. [4], [5]), or the dependence structure across attributes (e.g. [2], [6]). Here, this paper is taking a different approach by considering the suitable parameterization of randomization matrices that can lift the above-mentioned difficulties. In the absence of clear guidelines in the literature on how to parameterize those matrices in practice, we develop new theoretical results that allow us to characterize a simple yet intuitive structure for randomization matrices where basically a single parameter, to be evaluated by the data controller according to her information/privacy trade-off, allows to build randomization matrices. Then, it follows that this specific class of matrices exhibits simple properties in terms of inversion operation, that we also characterize. These properties imply a computationally cheap way for retrieving the inverse of randomization matrices. Finally, from these elements we develop a protocol called $\lambda$-randomization, a computationally low-cost, local or centralized multi-dimensional randomized response protocol that requires only three basic elements: a set of parameters ranging between 0 and 1 (one per attribute), the identity matrix, and the all-ones vector.

The remainder of this paper is organized as follows. Section 2 gives some background elements, needed later. Using recently proposed theoretical results, Section 3 develops a simple approach for the relationship between univariate randomization and multivariate randomization. Then, Section 4 presents some results on the parameterization of randomization matrices. All these elements are gathered into Section 5, that develops the $\lambda$-randomization protocol. Finally, Section 6 presents a small empirical illustration, while conclusions and future research directions are gathered in Section 7.

## 2  Background elements

### 2.1  Randomized response

Let $X$ denotes an original categorical attribute with $1, \ldots, r$ categories, and $Y$ its anonymized version. Given a value $X = u$, RR [7] computes a value $Y = v$ by using an $r \times r$ Markov transition matrix:

$$P = \begin{pmatrix} p_{11} & \cdots & p_{1r} \\ \vdots & \ddots & \vdots \\ p_{r1} & \cdots & p_{rr} \end{pmatrix} \qquad (1)$$

where $p_{uv} = \Pr(Y = v | X = u)$ denotes the probability that the original response $u$ in $X$ is reported as $v$ in $Y$, for $u, v \in \{1, \ldots, r\}$. To be a proper Markov transition matrix, it must hold that $\sum_{v=1}^{r} p_{uv} = 1 \ \forall u = 1, \ldots, r$. $P$ is thus *right stochastic*, meaning that any original category must be spread along the anonymized categories.

The usual setting in RR is that each subject computes her randomized response $Y$ to be reported instead of her true response $X$. This is called the *ex-ante* or local anonymization mode. Nevertheless, it is also possible for a (trusted) data curator to gather the original responses from the subjects and randomize them in a centralized way. This *ex-post* mode corresponds to PRAM [3]. Apart from who performs the anonymization, RR and PRAM operate the same way and make use of the same matrix $P$.

Let $\pi_1, \ldots, \pi_r$ be the proportions of respondents whose true values fall in each of the $r$ categories of $X$; let $\theta_v = \sum_{u=1}^{r} p_{uv} \pi_u$ for $v=1,\ldots,r$ be the probability of the reported value $Y$ being $v$. If we define by $\theta = (\theta_1, \ldots, \theta_r)^T$ and $\pi = (\pi_1, \ldots, \pi_r)^T$, then we have $\theta = P^T \pi$. Furthermore, if $P$ is nonsingular, it is proven in [7] that an unbiased estimator $\hat{\pi}$ of $\pi$ can be obtained as:

$$\hat{\pi} = (\widehat{P^T})^{-1} \hat{\theta} \qquad (2)$$

Thus, estimates of frequencies can be retrieved from the protected data set. Note that this procedure does not entail any privacy risk as only some estimates of the frequencies are retrieved, not specific responses that can be traced back to any individual.

Here, two sources of errors can plague the estimates. First, the $(\widehat{P^T})^{-1}$ part: even if $P$ is nonsingular, its conditioning can make it sensitive to numerical operations, in particular inversion. Indeed, if the matrix happens to be ill-conditioned, its inverse may only be a poor estimate of the real inverse [8]. Such issue makes the parameterization of randomization matrix a critical one.

Second, the $\hat{\theta}$ part: as $\theta$ is only an observed instance of the real proportions in the population, if it deviates from the expected values then additional errors are introduced in *Equation (2)*. In this paper, due to space constraints, and bearing in mind that the second type of error is a pure statistical and sampling issue, we will focus on the first possible type of errors, more related to privacy and how randomization is performed, and will discard the second.

RR is based on an implicit privacy guarantee called *plausible deniability* [9]. It equips the individuals with the ability to deny, with variable strength according to the parameterization of $P$, that they have reported a specific value. In fact, the more similar the probabilities in $P$, the higher the deniability. In the case where the probabilities within each column of $P$ are identical, it can be proved that perfect secrecy, or *perfect* *privacy*, in the Shannon sense is reached [10]: observing the anonymized attribute *Y* gives no information at all on the real value *X*. Under such parameterization of *P*, a privacy breach cannot originate from the release of an anonymized data set, as the release does not bring any information that could be used for an attack. However, the price to pay in terms of data utility is high [9].

### 2.2 Bistochastic privacy

In the rest of this paper, we will assume that the randomized response matrix *P* above fulfills the additional left stochasticity constraints that $\sum_{u=1}^{r} p_{uv} = 1 \ \forall v = 1, \ldots, r$. This makes *P bistochastic* (left stochasticity implying that any anonymized categories must come from the original categories). Moreover, it will also be assumed that $p_{uv} > 0 \ \forall u, v$: the transition matrix *P* has only strictly positive entries, meaning that any individual in any of the *r* categories can be reported in the anonymized attribute in any other of the *r* categories. As some of the transition probabilities can be made as small as desired, this is not really binding. However, this additional constraint makes *P* the transition matrix of an ergodic Markov chain [11]. In turn, that implies that *P* has a *unique* stationary distribution, which, as *P* is bistochastic, is the uniform distribution [11]. This last assumption ensures that protection will always range between no protection at all and perfect privacy (see below).

At first sight, one could wonder about the necessity of imposing an additional constraint on RR and its ex-post version PRAM, some well-established approaches for anonymization that have proved their merits over the years. However, it happens that the bistochasticity assumption connects several fields of the privacy literature, including the two most popular models, *k*-anonymity and ε-differential privacy, but also any Statistical Disclosure Control (SDC) method. We refer the reader to [12] for a detailed presentation of these results.

Beyond its unifying properties, the bistochastic version of RR offers additional advantages by clarifying and operationalizing the trade-off between protection and utility. Indeed, it is well-known that a bistochastic matrix never decreases uncertainty and is the only class of matrices to do so [13]. Stated otherwise, when a bistochastic *P* is applied to an original attribute X, its anonymized version Y will always contain more entropy than X. Remark that when *P* is only right stochastic, as in the traditional approach to RR, no particular relationship emerges.

Now, and as a direct consequence of this last property, the strength of anonymization (and equivalently the strength of plausible deniability), can be measured, as in cryptography for the strength of security, in terms of bits, through $H(P)$, the entropy rate of *P* (this rate being the average of the entropies of each row of *P* [14]). In the case of perfect privacy where all probabilities in *P* are equal, and that we will denote hereafter by $P^*$, we have $H(P^*) = \log_2 r$, which is the maximum achievable entropy for an $r \times r$ bistochastic matrix. Thus, after the choice of a suitable parameterization, the number of bits that *P* contains establishes a metric in terms of plausible deniability.

From these results, the definition of bistochastic privacy follows. We provide here the univariate case, where we seek to anonymize only one attribute to prevent disclosure (other definitions at the data set level can be found in [12]):

***Definition 1 (Univariate bistochastic privacy)***: *The anonymized version Y of an original attribute X is β-bistochastically private for 0 ≤ β ≤ 1 if:*
  i)     $Y = P'X$ with P bistochastic
  ii)    $\frac{H(P)}{H(P^*)} \geq \beta$.

An anonymized attribute satisfies $\beta$-bistochastic privacy if it is the product of a bistochastic matrix $P$ and the original attribute, and if the entropy rate of $P$ is at least $100\beta\%$ of the maximum achievable entropy. $H(P^*)$ represents the maximum "spending" that can be allocated to privacy, $\log_2 r$ bits. Thus, when $\beta = 1$, all the bits have been spent and the attribute has been infused with the maximum possible amount of uncertainty; in this case, perfect secrecy is achieved and it is clear that $Y = P^{*'}X$ returns the uniform distribution, breaking the link with all other attributes in the dataset and entailing a huge information cost. The other extreme case $\beta = 0$ means that the attribute has been left untouched, and no uncertainty has been injected, i.e. $H(P) = 0$. In that case, the data user gets the highest possible utility from the data. As a result, for $0 < \beta < 1$ there lies a continuum of cases where varying amount of uncertainty bits can be injected, which will guarantee a varying amount of protection and information. Unlike other privacy models, bistochastic privacy makes the trade-off between privacy and information explicit.

## 3     From univariate to multivariate randomized response

Consider $n$ individuals $i = 1, \ldots, n$ each holding one record $x_i = (x_{i,1}, \ldots, x_{i,m})$ that contains the values for $m$ categorical attributes (the case of numerical attributes will be tackled later in this paper). A data controller wants to perform data analysis on the pooled data of the $n$ individuals. However, the controller is not trusted such that none of the $n$ individuals want to disclose her true record, but only a protected version of it. In that setting, RR seems to be suited to the task.

A naïve protocol is for the controller to generate $m$ bistochastic matrices $P_1,\ldots,P_m$ that the $n$ individuals will use to perform RR locally and separately for each attribute. Let us illustrate this basic approach for $m=2$ binary attributes. The first attribute is randomized with the bistochastic matrix $\begin{pmatrix} 0.9 & 0.1 \\ 0.1 & 0.9 \end{pmatrix}$ and the second with $\begin{pmatrix} 0.7 & 0.3 \\ 0.3 & 0.7 \end{pmatrix}$. Once the individuals are done randomizing, they communicate their randomized values for the two binary attributes to the controller. In turn, using *Equation (2)* the controller can estimate the *marginal* distributions of the two attributes, providing that the two matrices are invertible (which is the case in this example).

This protocol has its merits but is naïve because, apart from the case where the two attributes are independent, it does not provide an estimate of the *joint* distribution. Indeed, only if the attributes are independent can the joint distribution be estimated from the marginals.

To abstract from the independence assumption, one way to proceed is to perform RR on the joint distribution, as suggested in [2], i.e. computing the cartesian products of the values that the categorical attributes have and then performing RR locally on this product. In the current example, if $m_1$ and $m_2$ can take the values $(a_1, a_2)$ and $(b_1, b_2)$, respectively, RR will be performed on the joint categorization $a_1 \times b_1$, $a_1 \times b_2$, $a_2 \times b_1$, $a_2 \times b_2$ using a bistochastic matrix of size 4. Then, after the individuals report their randomized values to the controller, she can estimate the joint distribution.

This approach is clearly more adequate in theory but suffers from one major drawback: combinatorial explosion. As the number of attributes grows, and/or if the number of categories for each attribute is large, this approach will clearly become intractable in practice. For instance, the computational cost of inverting the matrix in *Equation (2)* will turn unreasonable. Moreover, and even in the case of a data controller viewing the computational cost as acceptable, the conditioning of the matrix (i.e. its sensitivity to numerical operations, specifically inversion even if the matrix is theoretically invertible), may preclude to get a reliable enough estimate of the joint distribution [8]. Thus, in practice, RR is not immune to the curse of dimensionality.

In this paper, to avoid being too naïve while at the same time staying practical, we consider as a first step a new approach where randomization matrices are parameterized separately on each attribute but where, from the strength of randomization on each attribute, the data controller can control for the randomization of the joint distribution. We believe this is a suitable approach as parameterizing the randomization matrix directly for the joint distribution can be a daunting task given its generally large size. Moreover, it allows to control for what is happening to each attribute in terms of randomization, as some may necessitate varying degrees of protection. Finally, and as will be tackled later in this paper, it also allows to control for the degree of preservation of the covariances in the final, randomized data set.

To develop this new approach, we remark that the randomization of the joint distribution can be represented by the Kronecker product of the matrices used for the marginals. In the above example, it gives:

$$\begin{pmatrix} 0.9 & 0.1 \\ 0.1 & 0.9 \end{pmatrix} \otimes \begin{pmatrix} 0.7 & 0.3 \\ 0.3 & 0.7 \end{pmatrix} = \begin{pmatrix} 0.63 & 0.27 & 0.07 & 0.03 \\ 0.27 & 0.63 & 0.03 & 0.07 \\ 0.07 & 0.03 & 0.63 & 0.27 \\ 0.03 & 0.7 & 0.27 & 0.63 \end{pmatrix}$$

For instance, the first term of this matrix means that an individual with a true record *(a_1, b_1)* has a 0.63 probability of reporting to the controller, after randomization, the same record. Similarly, if that individual has a true record *(a_1, b_2)*, it has also a 0.63 probability of reporting the same record to the controller, as indicated by the second diagonal term. Now, considering the second probability in the first column, an individual with a record *(a_1, b_1)* has a 0.27 probability to report *(a_1, b_2)*; her record has been modified by randomized response.

Remark that the cartesian products of all possible attributes' values and the Kronecker product of the matrices used for the marginals lead to the same set of all possible combinations. However, the former is a direct combinatorial description not producing a matrix of the joint probabilities (thus needed to be subsequently parameterized), while the latter, using this description, produces a matrix of the joint randomization probabilities based on the randomization matrices used for each attribute.

Moreover, recalling the following Theorem, this Kronecker product is also bistochastic:

***Theorem 1 (Marshall, Olkin, and Arnold [13]):*** *If P and Q are m×m and n×n bistochastic matrices, respectively, then the Kronecker product P ⊗ Q is an mn × mn bistochastic matrix.*

Thus, the Kronecker product can be interpreted in the same way as any bistochastic matrices used for RR. For instance, its diagonal contains the probability that the randomized records are the true ones, meaning that the diagonal values will indicate a certain level of "truthfulness" at the record level, similarly to RR applied to a single attribute (see *Equation (1)*).

Now, by the associative property of the Kronecker product, and using repeatedly *Theorem 1*, the Kronecker product of an arbitrarily finite number of bistochastic matrices will also be bistochastic. We can thus repeat this procedure as the number of attributes grows. Next, we recall a recent result on bistochastic matrices:

***Theorem 2 (Ruiz [15]):*** *If $P_1,...,P_m$ are $n_1×n_1, ..., n_m×n_m$ bistochastic matrices with entropy rates $H(P_1),...,H(P_m)$, respectively, then the entropy rate of the Kronecker product $P_1 \otimes ... \otimes P_m$ is the sum of the entropy rates of each matrix composing the product:*

$$H(\otimes_{p=1}^m P_p) = \sum_{p=1}^m H(P_p) \qquad (3)$$

Applied to multi-dimensional RR, *Theorem 2* states that the level of randomization (as convey by the entropy rates of bistochastic matrices) of the joint distribution *relates additively* to the randomization of each attribute. As a result, a simple and rather expected relationship emerges for the randomization of a data set with an arbitrary number of attributes: *the more (less) each attribute is randomized, the more (less) the joint distribution is randomized*.

Obviously, this simple, new approach is partially satisfactory. While to get a grasp on the level of randomization of a joint distribution a data controller can simply compute the sum of the entropy rates for each matrix applied to each attribute, she can only go that far, but she will still be left with the same practical problem as before. As the number of attributes increases, the resulting Kronecker product will become too huge to be tractable, especially to be inverted easily. Stated otherwise, the randomization property of this product can be observed, but the product in itself cannot be easily manipulated when too large. Moreover, this approach is blind as to which relationships across the attributes are preserved to a lesser or greater extent. However, it provides a starting point. As will be developed in the next section, by using this approach the two problems above-mentioned can be circumvented with specific, intuitive parameterization of bistochastic matrices.

# 4  Results on the parameterization of bistochastic matrices

The parameterization of Markov matrices (being bistochastic or not) for RR is a surprisingly open problem in the privacy literature. Obviously, the more the probability masses are spread off-diagonal, the stronger is the randomization (and thus the higher is plausible deniability), and thus the stronger is the individuals' protection. But beyond this simple fact, we have not been able to identify further guidelines on their parameterizations in the literature. As mentioned above, Bistochastic Privacy establishes entropy as a useful metric to measure the strength of randomization [12]. However, it does not tell how to concretely parameterize bistochastic matrices to reach a given level of entropy, while it is known that different parameterizations can lead to similar levels of entropy [14].

In this section, we aim to advance on this issue. By developing new results on bistochastic matrices, we seek to provide guidelines as to their parameterization for privacy and RR, and to establish more intuitiveness to interpret their structures. We start by reminding a classic theorem on bistochastic matrices:

***Theorem 3 (Birkhoff-Von Neumann [13])***: *If an r×r matrix P is bistochastic, then there exist $\lambda_1, \ldots, \lambda_J \geq 0$ with $\sum_{j=1}^{J} \lambda_j = 1$ and $P_1, \ldots, P_J$ permutation matrices such that:*

$$P = \sum_{j=1}^{J} \lambda_j P_j \qquad (4)$$

*Theorem 3* (often called the Birkhoff-Von Neumann decomposition) states that any bistochastic matrix can always be expressed as a convex combination of permutation matrices. Note that this combination may not be unique [13]. Next, we establish a corollary to this theorem which, to the best of our knowledge, has never been characterized before:

***Corollary 1***: *If an r×r matrix P is bistochastic with $P_{ij}>0$ $\forall i,j$, then there always exists an admissible Birkhoff-Von Neumann decomposition in which the identity matrix I appears with positive weight $0< \lambda \leq 1$:*

$$P = \lambda I + \sum_{j=1}^{J}(1-\lambda)\mu_j P_j \qquad (5)$$

***Proof***: *see Appendix*

*Corollary 1* establishes that one can *always* decompose any bistochastic matrices with strictly positive terms (i.e. the ergodic case assumed in this paper; see above) as a convex combination of the identity matrix and some permutation matrices. The goal of this result is to isolate, from a privacy perspective, the randomization case where, in fact, no randomization is performed. Individuals report their true values to the data controller if all the weight is put on *I* in *Equation (5)*. More generally, the probability attached to this peculiar setting, $\lambda$, is driven by the diagonal terms, which we remind indicates the level of truthfulness in RR.

As *Equation (5)* is a Birkhoff-Von Neumann decomposition, it may not be unique. Exploring admissible decompositions is beyond the scope of this paper, but among them we will retain a specific one, that is when the permutation matrices in *Equation (5)* are the average over all possible permutations. To pinpoint this case, let $P_\pi$ be the permutation matrix associated to a permutation $\pi \in S_\pi$, where $S_\pi$ denotes the group consisting of all possible reorderings of the set $\{1, \ldots, J\}$. For a fixed pair $(i, j)$ there are exactly $(J - 1)!$ permutations $\pi$ with $\pi(i) = j$. Hence, we have:

$$\sum_{\pi \in S_\pi} (P_\pi)_{ij} = (J - 1)!$$

Dividing by the number of all possible permutations $J!$, one gets:

$$\frac{1}{J!} \sum_{\pi \in S_\pi} (P_\pi)_{ij} = \frac{1}{J}$$

That is, the convex combination consisting of averaging all possible permutations matrices is an admissible Birkhoff-Von Neumann decomposition of the perfect privacy matrix $P^*$ mentioned above, where all probabilities are equal. We can thus use it in *Equation (5)* as a choice for $\sum_{j=1}^{J}(1 - \lambda)\mu_j P_j$. Then, we can rewrite *Equation (5)* as:

$$P = \lambda I + (1 - \lambda)P^* \text{ with } 0 < \lambda \leq 1 \qquad (6)$$

Naturally, by enforcing $\sum_{j=1}^{J}(1 - \lambda)\mu_j P_j$ in *Equation (5)* to be the average of all possible permutation matrices, we are constraining the structure of the type of matrix $P$ that can be decomposed using *Equation (6)*. In fact, such a bistochastic matrix will have to be symmetric, with all terms on-diagonal equal and all terms off-diagonal equal. However, what is lost in freedom of parameterization is, we believe, offset by several features. The first one is that it eases considerably the task of the data controller when it comes to parameterizing the randomization matrices. Under *Equation (6)*, the controller has simply to consider two polar situations, no randomization at all (the identity matrix) or maximum randomization (the perfect privacy matrix), and to decide through a single parameter $\lambda$ how to weigh them. The randomization matrix then follows. Arguably, it is a very simple and intuitive way of deciding how randomization should be conducted, and also a straightforward procedure to generate randomization matrices.

A second feature is that the class of matrices $P$ satisfying the decomposition of *Equation (6)* (that we will denote hereafter by $P_{(\lambda)}$) happens to fulfill special properties in term of invertibility as their inverses, as well the inverse of the Kronecker product of many of such matrices, can be computed exactly, in fact without the need of numerically inverting any matrices. As such, one does not need to bother with the conditioning of the matrices, or even their sizes. We demonstrate this with the two results below.

***Property 1***: *If* $P = \lambda I + (1 - \lambda)P^*$, *with P of size N and* $0 < \lambda \leq 1$, *then its inverse is:*

$$P^{-1} = \frac{1}{\lambda}(I - P^*) + P^* \qquad (7)$$

***Proof***: see Appendix

***Property 2***: *Let $n_i > 1$ be positive integers and $\forall i = 1, \ldots, I$ let $u_i \in \mathbb{R}^{n_i}$ be the all-ones column vector. Define $P_i = \lambda_i I_{n_i} + (1 - \lambda_i) P^*_{n_i}$ for $0 < \lambda_i \leq 1$, and let $\varepsilon = (\varepsilon_1, \ldots, \varepsilon_I) \in \{0,1\}^I$ a binary indicator vector. For each coordinate, define:*

$$T_i(\varepsilon_i) = \begin{cases} I_{n_i} & \text{for } \varepsilon_i = 0 \\ u_i u_i^T & \text{for } \varepsilon_i = 1 \end{cases}$$

*Then, it holds that:*

$$(\otimes_{i=1}^I P_i)^{-1} = \sum_{\varepsilon \in \{0,1\}^I} \left[ \prod_{i=1}^I \left(\frac{1}{\lambda_i}\right)^{1-\varepsilon_i} \left(-\frac{1-\lambda_i}{\lambda_i n_i}\right)^{\varepsilon_i} \right] \otimes_{i=1}^I T_i(\varepsilon_i) \qquad (8)$$

***Proof***: see Appendix

These two properties show that, when performing RR with the $P_{(\lambda)}$ class of matrices derived above, *the data controller can abstract from any numerical and computational difficulties when trying to retrieve estimates of the joint distributions*. Following *Equation (2)*, the inverse of the matrices can be computed *exactly*, as the sum of basic elements: the $\lambda$'s chosen to parameterize the randomization matrices, the identity matrix, and the all-ones vector (and remark that $P^* = \frac{uu^T}{N}$, where $u \in \mathbb{R}^N$ is the all-ones vector). Matrix addition, albeit if here the matrices can be quite large, is one of the less computationally intensive matrix operations (especially when the matrices have a basic structure, such as the ones considered here). Stated otherwise, what is proposed in this paper is *a computationally cheap way of computing inverses of randomization matrices to estimate multivariate distributions, based on randomization matrices with an intuitive parameterization and structure*.

To conclude this section, we turn to the issue of the relationships between attributes. Applying RR separately on each attribute will invariably degrade their dependencies, except when the attributes are already independent, which will not be the case in most practical cases. Here, we will focus on the dependency between a pair of attributes $x$ and $y$, both of size $n$, of means $\bar{x}$ and $\bar{y}$, and randomized with two bistochastic matrices $A$ and $B$. Their centered versions are denoted $\tilde{x}$ and $\tilde{y}$ and their randomized versions by $x' = Ax$ and $y' = By$. Their original covariance is given by:

$$cov(x, y) = \frac{1}{n}(x - \bar{x}u_n)^T(y - \bar{y}u_n) = \frac{1}{n}\tilde{x}^T \tilde{y}$$

As $A$ and $B$ are bistochastic, $Au_n = Bu_n = u_n$. Thus, $\bar{x'} = x'$ and $\bar{y'} = y'$. After randomization, the covariance becomes:

$$cov(x', y') = \frac{1}{n}(Ax - \bar{x}u_n)^T(By - \bar{y}u_n) = \frac{1}{n}\tilde{x}^T(A^T B)\tilde{y}$$

The covariance is altered by the term $A^T B$. Now, assume that A and B are of the $P_{(\lambda)}$ class. As the attributes are of the same size, A and B now only differ from their parameters $\lambda_1$ and $\lambda_2$, and we have:

$$cov(x', y') = \frac{1}{n}\tilde{x}^T[(\lambda_1 I_n + (1 - \lambda_1)P^*_n)^T(\lambda_2 I_n + (1 - \lambda_2)P^*_n)]\tilde{y}$$

By the property of the transpose operation and as $I_n$ and $P_n^*$ are idempotent, the expression above simplifies to:

$$cov(x', y') = \frac{1}{n} \tilde{x}^T (\lambda_1 \lambda_2 I_n + (1 - \lambda_1 \lambda_2) P_n^*) \tilde{y} \qquad (9)$$

When using $P_{(\lambda)}$ matrices, the term altering the covariance after randomization has the same structure than the matrices used, except that the coefficient of the convex combination is the product of the $\lambda$'s. As a result, under this setting covariance will be weakly altered *if both* of the $\lambda$'s are close to 1 and strongly altered *if at least one* of the $\lambda$'s is close to zero. This is a rather intuitive result that falls back on the classic information/protection trade-off: the stronger is randomization, the lower is the preservation of dependencies between attributes. However, *Equation (9)* allows the data controller to parameterize the randomization matrices according to the relationships that she may seek to preserve to a greater or lesser extent in the randomized version of the dataset.

## 5 λ-randomization

Using the elements developed above, we can now formulate λ-randomization, a local or centralized protocol allowing to perform RR in a simple way. We keep using the setting of *n* individuals $i = 1, \ldots, n$, each holding one record $x_i = (x_{i,1}, \ldots, x_{i,m})$ for *m* categorical attributes. The protocol is the following:

***Protocol: λ-randomization***
1. The controller generates $(\lambda_1, \ldots, \lambda_m)$ with $0 < \lambda_j \leq 1 \, \forall j = 1, \ldots, m$. She does so according to her views about the strength of randomization to be applied on each attribute, and thus her appraisal about the privacy/information trade-off. She also considers which relationships should be preserved, or not, among pairs of attributes, following *Equation (9)*. Note that $0 < \lambda_j$ enforces a minimum level of truthfulness, even if very low.
2. The controller generates $P_{(1)}, \ldots, P_{(m)}$ randomization matrices using *Equation (6)*.
3. The controller computes the entropy rates of matrices $P_{(1)}, \ldots, P_{(m)}$. Summing those rates, by *Equation (3)* she gets an evaluation of the overall protection level of the data set. If that level is judged as unsatisfactory, $(\lambda_1, \ldots, \lambda_m)$ are re-generated to reach the desired protection level. If not, she computes the Kronecker product of these matrices.
4. Each individual $i = 1, \ldots, n$ performs RR with the matrix $P_{(1)} \otimes \ldots \otimes P_{(m)}$ and publishes the randomized record $x'_i = (x'_{i,1}, \ldots, x'_{i,m})$.
5. The controller gathers the distribution of randomized records. She computes the exact inverse of $P_{(1)} \otimes \ldots \otimes P_{(m)}$ using *Equation (8)* and then estimates the true distribution using *Equation (2)*.

Overall, this protocol entails low computational cost, avoiding the traditional pitfalls of multi-dimensional RR [2]. It also enforces, and at the same time clarifies, the parameterizations of randomization matrices, as an expression of what is the balance between protection (as convey by plausible deniability in RR) and information thought adequate by the data controller. As such, this approach is not attached to any privacy model but is directly based on the classical information/protection trade-off in privacy. However, it can be noted that the $P_{(\lambda)}$ class of matrices proposed in this paper happens to contain the matrix of an ε-differentially private randomized response scheme [16].

Finally, we consider the case where one or several numerical attributes are present by remarking that bistochastic matrices can be applied on *both* categorical and numerical attributes. In the categorical case, the original proportions of respondents whose values fall in each category will be changed. In the numerical case, *the individuals are used as categories,* and the resulting randomized numerical values are convex combinations of the original ones. That being said, in the local scenario of RR the data controller needs to know *ex-ante* upon which categorization randomization is going to take place, to be able to establish the size of the randomization matrices. In the case of a numerical attribute, except if it is preliminary categorized, the local scenario then cannot be run. Thus, λ-randomization can accommodate numerical attributes in the centralized scenario (i.e. PRAM), or the local scenario but only after having preliminary categorized the numerical attributes.

## 6 Empirical example

In the example below, we act as a data controller who generates bistochastic matrices to randomized three categorical attributes, each with 5 categories. The original proportions are simulated assuming *n=100* individuals. We select three scenarios for *λ*'s values: (0.9,0.5,0.4), (0.3,0.2,0.1), (0.6,0.7,0.4). Using the λ-randomization protocol, the results are shown in Table 1.

Table 1. Example under different λ's scenarios.

| λ's values | Randomization strength (as a % of the maximum possible strength) | | | |
|---|---|---|---|---|
| | Attribute 1 | Attribute 2 | Attribute 3 | Joint distribution |
| {0.9; 0.8; 0.7} | 25% | 41% | 56% | 31% |
| {0.3; 0.2; 0.1} | 92% | 96% | 99% | 72% |
| {0.6; 0.7; 0.4} | 68% | 52% | 85% | 51% |

In this example, results on the strength of randomization are expressed relative to the maximum possible strength (i.e. as a percentage of the perfect privacy case, as in bistochastic privacy). The first scenario depicts the case where relatively high weight is put on the identity matrix for each of the three attributes. In that case, the joint distribution is weakly protected, at slightly less than one third of the maximum possible strength. The second scenario displays the opposite case, where relatively high weight is applied to the perfect privacy case. The joint distribution is thus strongly protected, at almost three-fourth of the maximum randomization possible. Finally, the third scenario is the intermediate case, where the three attributes are mildly protected. In that scenario, the joint distribution is also mildly protected, at one-half of the maximum randomization strength possible.

Remark that in this small example, randomization could have been conducted locally or globally. In both cases the data controller, after having retrieved the randomized individual responses, can proceed with the rest of the protocol. In particular, and for example considering the third scenario, she can compute the inverse for the joint distribution. Using *Equation (8)*, each factor has two choices ($(I - P^*)$ or $P^*$), so the expansion has $2^3$ terms. We group them explicitly by how many times $(I - P^*)$ appears (with each matrix being of size 5x5):

(i) No $(I - P^*)$ terms: $P^* \otimes P^* \otimes P^*$

(ii) Exactly one $(I - P^*)$: $\frac{1}{0.6}(I - P^*) \otimes P^* \otimes P^* + \frac{1}{0.7} P^* \otimes (I - P^*) \otimes P^* + \frac{1}{0.4} P^* \otimes P^* \otimes (I - P^*)$

(iii) Exactly two $(I - P^*)$'s: $\frac{1}{0.6 \times 0.7}(I - P^*) \otimes (I - P^*) \otimes P^* + \frac{1}{0.6 \times 0.4}(I - P^*) \otimes P^* \otimes (I - P^*) + \frac{1}{0.7 \times 0.4} P^* \otimes (I - P^*) \otimes (I - P^*)$

(iv) Exactly three $(I - P^*)$'s: $\frac{1}{0.6 \times 0.7 \times 0.4}(I - P^*) \otimes (I - P^*) \otimes (I - P^*)$

By summing up the four terms (i) to (iv), one obtains $\left(P_{(1)} \otimes P_{(2)} \otimes P_{(3)}\right)^{-1}$, the exact and complete, closed-form inverse of the 125×125 matrix for the randomization of the joint distribution. It is written as a sum comprising only two types of terms: *(i)* the matrices $(I - P^*)$ and $P^*$; *(ii)* some scalar coefficients which are the inverse values of the coefficients used to parameterize the randomization matrix of each attribute.

# 7 Conclusions and future work

This paper considered multi-dimensional response, an appealing anonymization technique offering rigorous privacy guarantees while allowing, in principle, to retrieve estimates of the true distributions. However, as with many anonymization techniques, it is not immune to the curse of dimensionality. To tackle this issue, and based on new theoretical results, this paper introduced a specific class of randomization matrices carrying an intuitive interpretation regarding the information/privacy trade-off. Matrices in this class also happen to be computationally cheap to invert.

From these elements and using a simple relationship between univariate and multivariate randomization that we also introduced, we proposed λ-randomization, a multi-dimensional randomized response protocol based on solely three ingredients: a set of parameters ranging between 0 and 1 (one per attribute), the identity matrix, and the all-ones vector. From these, multi-dimensional response can be conducted in a simple fashion, locally or globally, and with small practical difficulties.

This paper opens several lines for future research. One of them is to introduce in λ-randomization errors attached to frequencies' estimates, which as we mentioned is more a statistical and sampling issue than a privacy one. However, it remains an important problem in randomized response. Another avenue for research is to conduct further empirical work on real-life data. Finally, another path is to derive rules, within the present approach and for the local scenario, yielding good estimates for numerical attributes.

# Appendix

### Proof of Corollary 1:

Since every entry in P is strictly positive, in particular every diagonal entry is strictly positive. Define $\lambda = \min_{1 \leq i \leq r} P_{ii}$. Then we have $\lambda > 0$. Now, define the matrix $R = \frac{P - \lambda I}{1 - \lambda}$.
We check that R is bistochastic:
- For $i \neq j$ we have $R_{ij} = \frac{P_{ij}}{1-\lambda} > 0$. For $i = j$ we have $R_{ii} = \frac{P_{ii}-\lambda}{1-\lambda} \geq 0$ as by definition $\lambda \leq P_{ii}$. Thus, $R_{ij} \geq 0 \; \forall i, j$.
- For any row i, $\sum_{j=1}^{r} R_{ij} = \frac{1}{1-\lambda}\sum_{j=1}^{r}(P_{ij} - \lambda \delta_{ij}) = \frac{1}{1-\lambda}(1 - \lambda) = 1$ (with $\delta_{ij}$ the Kronecker delta equal to 1 if i=j, 0 otherwise). This holds because P is bistochastic by assumption, thus $\sum_{j=1}^{r} P_{ij} = 1$ and $\sum_{j=1}^{r} \lambda \delta_{ij} = 1$
- For any column j, through the same reasoning $\sum_{i=1}^{r} R_{ij} = \frac{1}{1-\lambda}\sum_{i=1}^{r}(P_{ij} - \lambda \delta_{ij}) = \frac{1}{1-\lambda}(1 - \lambda) = 1$

R is indeed bistochastic: its terms are nonnegative, and each of its row and column sums to one. Thus, we can apply the Birkhoff-Von Neumann decomposition to it, with $\mu_1, \ldots, \mu_J \geq 0$ and $\sum_{j=1}^{J} \mu_j = 1$:

$$R = \sum_{j=1}^{J} \mu_j P_j$$

Substituting the definition of R gives:

$$\frac{P - \lambda I}{1 - \lambda} = \sum_{j=1}^{J} \mu_j P_j$$

and we have:

$$P = \lambda I + (1 - \lambda)\sum_{j=1}^{J} \mu_j P_j = \lambda I + \sum_{j=1}^{J}(1 - \lambda)\mu_j P_j$$

The coefficients $\lambda$ and $(1 - \lambda)\mu_j$ are nonnegative and sum to $\lambda + (1 - \lambda)\sum_{j=1}^{J} \mu_j = 1$. Because $\lambda > 0$, the identity matrix I appears with strictly positive weight in the Birkhoff-Von Neumann decomposition of Equation (5).

***Proof of Property 1:***

Remark first that $P^* = \frac{uu^T}{N}$, where $u \in \mathbb{R}^N$ is the all-ones column vector. We can thus write P as a scalar multiplied by a rank-one perturbation of the identity matrix:

$$P = \lambda(I + \alpha uu^T) \text{ with } \alpha = \frac{1-\lambda}{\lambda N}$$

Hence

$$P^{-1} = \lambda^{-1}(I + \alpha uu^T)^{-1}$$

We can here apply the Sherman-Morrison formula [17], with special cases A=I and v=u, which gives:

$$(I + \alpha uu^T)^{-1} = I - \frac{\alpha uu^T}{1 + \alpha u^T u}$$

provided that $1 + \alpha u^T u \neq 0$. Here, $u^T u = N$ and we have $1 + \alpha u^T u = \frac{1}{\lambda}$, therefore $1 + \alpha u^T u \neq 0$ by assumption.

We then have:

$$\frac{\alpha}{1 + \alpha u^T u} = \alpha\lambda = \frac{1-\lambda}{N}$$

Substituting back:

$$(I + \alpha uu^T)^{-1} = I - \frac{1-\lambda}{N} uu^T$$

Finally, after rearrangement we get:

$$P^{-1} = \frac{1}{\lambda}(I - P^*) + P^*$$

Remark that P is always invertible when $\lambda \neq 0$.

***Proof of Property 2:***

From Property 1, we know that $P_i^{-1} = \frac{1}{\lambda_i}\left(I_{n_i} - P_{n_i}^*\right) + P_{n_i}^*$. A Kronecker product is invertible if and only if all its terms are invertible, which is the case here by assumption as $\lambda_i \neq 0$. Then, by property of the Kronecker product, we have:

$$(\otimes_{i=1}^I P_i)^{-1} = \otimes_{i=1}^I P_i^{-1}$$

This last equation is already sufficient to show that, in conjunction with Property 1, computing the exact inverse of the Kronecker product does not require any matrix inversion operation per se. However, this equation can be expended using the distributivity of the Kronecker product on $\lambda_i I_{n_i} + (1-\lambda_i)P_{n_i}^*$. After manipulations and rearrangements, we get:

$$\otimes_{i=1}^I \left[\frac{1}{\lambda_i}\left(I_{n_i} - P_{n_i}^*\right) + P_{n_i}^*\right]^{-1} = \sum_{\varepsilon \in \{0,1\}^I} \left[\prod_{i=1}^I \left(\frac{1}{\lambda_i}\right)^{1-\varepsilon_i}\left(-\frac{1-\lambda_i}{\lambda_i n_i}\right)^{\varepsilon_i}\right] \otimes_{i=1}^I T_i(\varepsilon_i)$$